\documentclass[final]{svjour2}
\smartqed 
\usepackage{graphicx}
 \usepackage{theorem}
\theorembodyfont{\upshape}
\newtheorem{teo}{Theorem}[section]
\newtheorem{pro}{Proposition}[section]

\usepackage{epsfig}
\usepackage{lscape}
\usepackage{epsf}
\usepackage{amsmath,amssymb}
\usepackage{array}
\usepackage{color}
\begin{document}




\title{On the number of clusters for planar graphs}

\author{J.-M. Billiot \and F. Corset \and E. Fontenas}

\institute {LabSAD, EA 3698, BSHM, Universit\'e Pierre Mend\`es France, 1251 Avenue Centrale, BP 47, 38040 Grenoble Cedex 9, France\\
\email {Jean-Michel.billiot@upmf-grenoble.fr}}

\maketitle

\begin{abstract}
The Tutte polynomial is a powerfull analytic tool to study the structure of planar graphs. In this paper, we establish some relations between the number of clusters per bond for planar graph and its dual : these relations bring into play the coordination number of the graphs. The factorial moment measure of the number of clusters per bond are given using the derivative of the Tutte polynomial. Examples are presented for simple planar graph. The cases of square, triangular, honeycomb, Archimedean and Laves lattices are discussed.
\end{abstract}

\section{Introduction}

It is well known that the Tutte polynomial is a powerfull tool in order to explore phase transition via the Lee and Yang theorem. It allows the exact calculation of the partition function using Fortuin-Kasteleyn representation of the Potts model for a large family of graphs. This representation shows a  strong link between phase transition and percolation purposes.
 We recall the famous paper \cite{Sykes64} on exact critical values for site and bond percolation using matching and dual graph and the number of cluster per site or per bond. A result of Kesten is proving that bond percolation thresholds of a dual pair of periodic planar lattices sum 1 and then that the exact  bond percolation threshold of the self dual square lattice is 0.5 \cite{Kesten80}. Wierman \cite{Wierman81} establishes exact bond percolation threshold for the triangular and honeycomb lattice and \cite{Wierman02} makes recent progress on bond percolation on Archimedean and Laves lattices.
As the chromatic polynomial is a particular case of the Tutte polynomial, since the proof of the four colour theorem for planar graph \cite{Appel77a,Appel77b}, a large number of conjectures have been proposed on the real, integer and complex zeros of the chromatic polynomial for different class of graphs, see \cite{Jackson03}. In fact, Sokal \cite{Sokal04} prove that complex zeros of chromatic and Tutte polynomial are dense in the complex plane. A large class of graphs has been studied : for graphs of bounded degree we can see the work of Procacci et al. \cite{Procacci04} and for graphs and matroids determined by their Tutte polynomials see \cite{Mier03} and references therein.

The paper is organized as follows. First, we recall the definition of Tutte polynomial of a graph and its dual. In a second part, some relations between the number of clusters per bond for planar graph and its dual are established using the coordination number. The factorial moment measure of the number of clusters per bond are given using the derivative of the Tutte polynomial. The cases of self-dual graphs, square, triangular, honeycomb, Archimedean and Laves lattices are discussed in the end of this paper.

\section{Tutte Polynomial}
\subsection{Definition}
Let be $G=(V,E)$ an undirected planar graph, where $V$ is the set of vertices and $E$ is the set of edges. The main theorem of this paper is based on the Tutte polynomial of a graph G. Given a connected graph $G$, the Tutte polynomial denoted by
 $T(G,x,y)$ can be obtained by computing the four following rules :\\
(1) If $G$ has no edges, then $T(G,x,y)=1$.\\
(2) If $e$ is an edge of $G$ that is neither a loop nor an isthmus, then $$T(G,x,y)=T(G'_e,x,y)+T(G''_e,x,y)$$ where $G'_e$ is the graph $G$ with the edge $e$ deleted and $G''_e$ is the graph $G$ with the edge $e$ contracted.\\
(3) If $e$ is an isthmus, then $T(G,x,y)=xT(G'_e,x,y)$\\
(4) If $e$ is a loop, then $T(G,x,y)=yT(G''_e,x,y)$.\\

\noindent{\bf Example :} The Tutte polynomial associated with a cycle of length $n$, $C_n$,
is given by $$T(C_n, x,y)=\sum_{k=1}^{n-1}x^k+y.$$

\subsection{Dual graph}
Given a connected planar graph $G$ with $n$ vertices and $m$ edges,
we assume that this graph has $f$ faces. The Euler relation gives
the following equality~: $$f=m-n+\alpha$$ with $\alpha$ is Euler-Poincar\'e caracteristic. In this paper, we are interested in planar lattice, which corresponds to $\alpha=1$. We can define the dual graph $G^{\star}$ of $G$ :
for this, it's necessary to add an exterior virtual face to the graph $G$. Thus, each vertice of $G^{\star}$ corresponds to a face of $G$ and two vertices of $G^{\star}$
are connected by an edge if the corresponding faces in $G$ have a boundary edge in common in $G$. We can write
$n^{\star}=f+1$, $m^{\star}=m$ where $n^{\star}$ is the number of
vertices of $G^{\star}$ and $m^{\star}$ the number of edges of
$G^{\star}$.

Then, we have the following relation between the two Tutte polynomial $T(G,x,y)=T(G^{\star},y,x)$.
For example, for self-dual graphs that means $G=G^{\star}$, the Tutte polynomial is symmetric in
$x$ and $y$. But, we don't have reciprocal result.
In fact, for example we can find different planar graphs having the same Tutte polynomial (see Figure 1 and \cite{Mier03} for details).

\begin{figure}[h]
\centering
\begin{tabular}{cc}
\includegraphics[width=4cm]{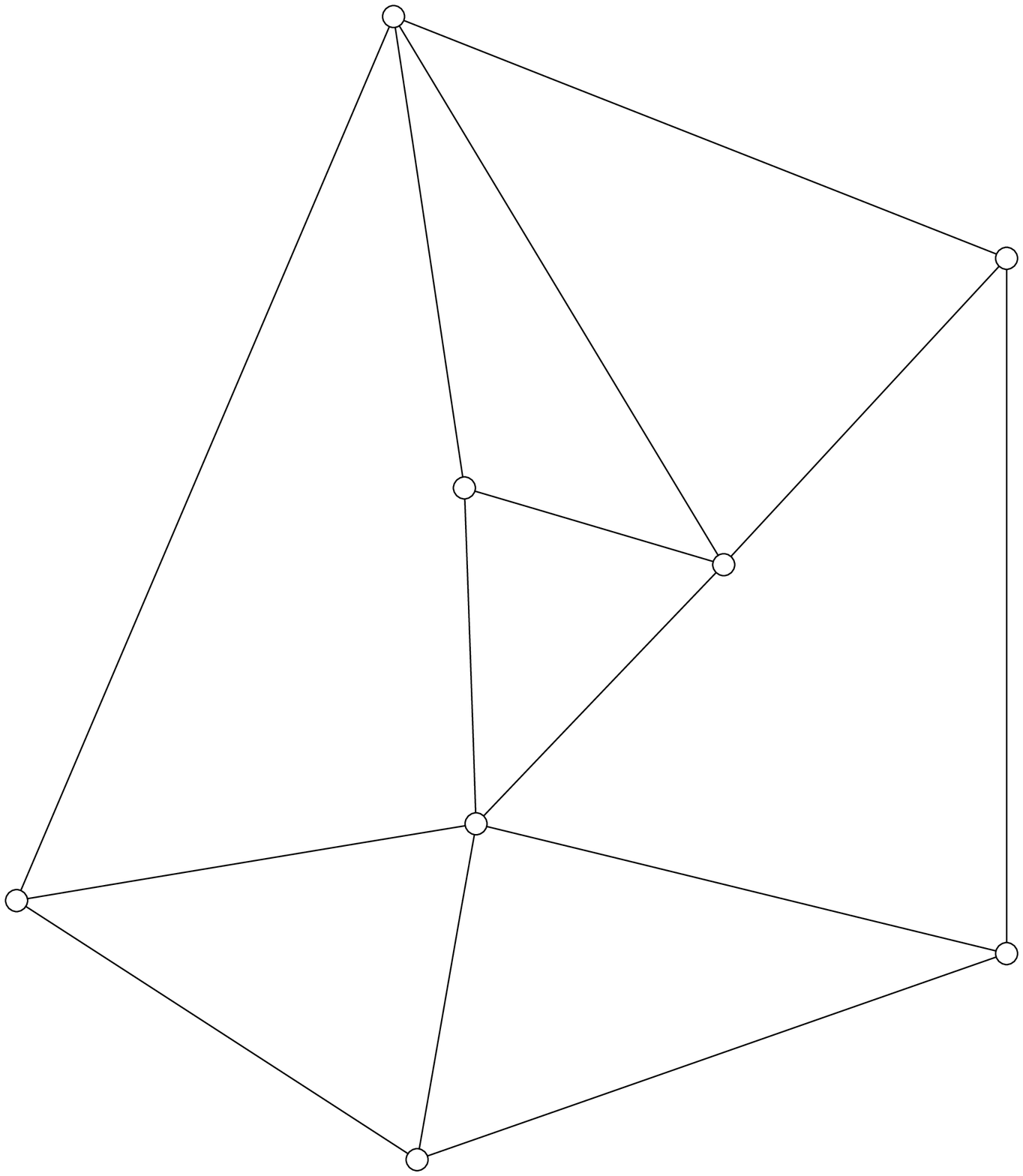} & \includegraphics[width=4cm]{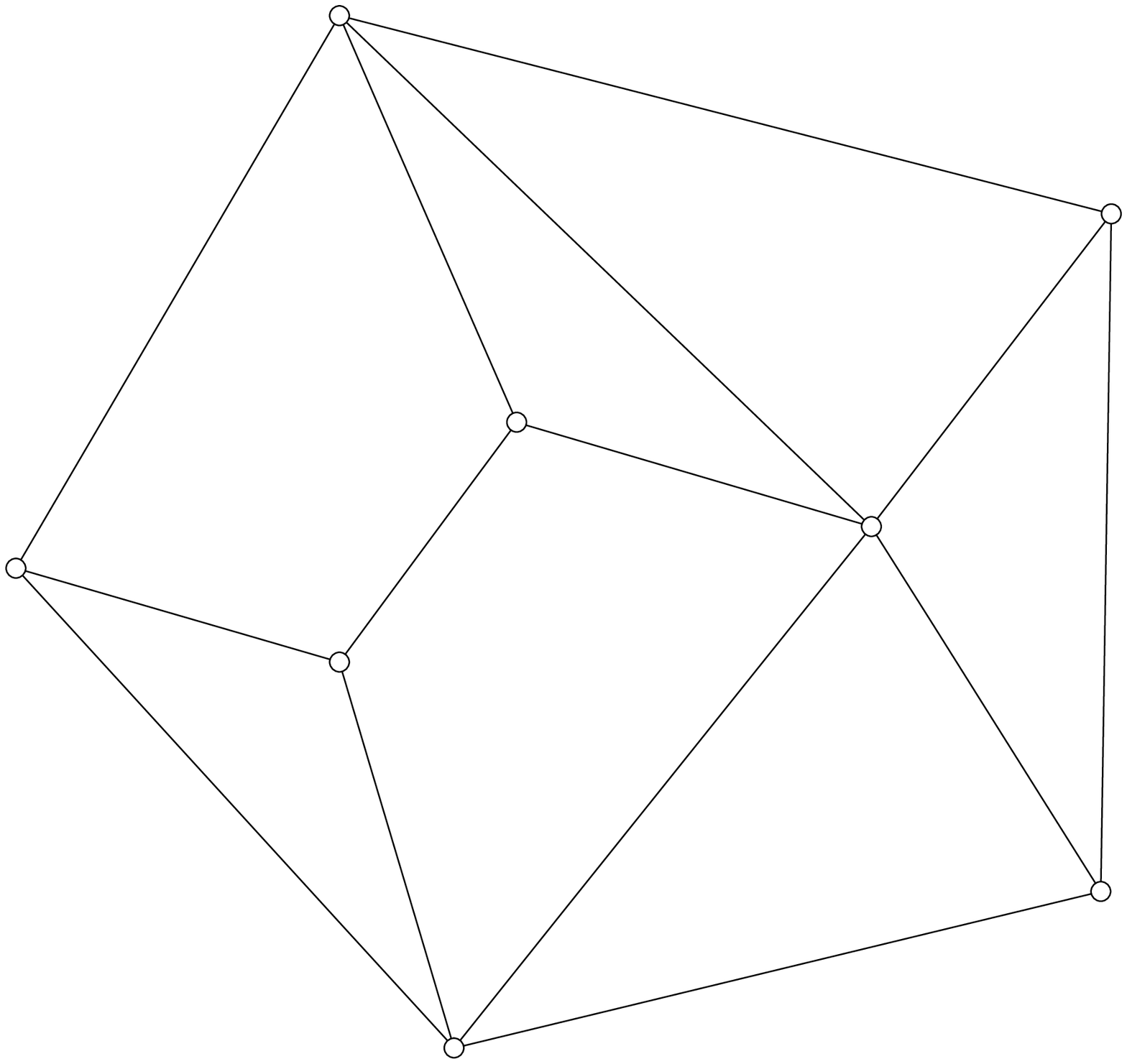}\\
\end{tabular}
\caption{\footnotesize two  graphs with the same Tutte polynomial.}

\end{figure}

\noindent{\bf An other representation of the Tutte polynomial}\\
We can introduce in the Tutte polynomial the notion of rank function
$r$ : the rank function of a graph $G$ is a function on the subsets
of $E$, which associates to each $A\subset
E$ the value $r(A)=|V|-k(G,A)$ where $V$ denotes the set of
vertices and $k(G,A)$ is the number of connected components of the
graph with vertex set $V$ and edge set $A\subset E$. This
definition provides an explicit formula for the Tutte polynomial :
$$T(G,x,y)=\sum_{A\subset E}(x-1)^{r(E)-r(A)}(y-1)^{|A|-r(A)}.$$
In this paper, we just consider connected graph : in this case, if
we set $n=|V|$, $m=|E|$, the Tutte polynomial may be written
as \begin{eqnarray*}T(G,x,y)&=&\sum_{A\subset
E}(x-1)^{k(G,A)-1}(y-1)^{k(G,A)+|A|-n}\\
&=&\frac1{(x-1)(y-1)^n}\sum_{A\subset
E}(y-1)^{|A|}[(x-1)(y-1)]^{k(G,A)}.\end{eqnarray*}

\noindent{\bf Two important values of the Tutte polynomial}\\
(1) If we calculate this Tutte polynomial on the curve
$(x-1)(y-1)=1$ or more specially for $x=1/p$ and $y=1/(1-p)$, we
obtain
$$T(G,1/p,1/(1-p)=\frac{1}{(x-1)(y-1)^n}\sum_{k=0}^mC_m^k(y-1)^k=\frac{y^m}{(y-1)^n(x-1)}$$
which gives
\begin{equation}\label{p(1-p)}
T(G,1/p,1/(1-p)=\frac{1}{(1-p)^{m-n+1}p^{n-1}}.
\end{equation}
(2) Now, we look at the curve $(x-1)(y-1)=q$, where $q$ is a real.
For $x=1+q(1-p)/p$ and $y=1/(1-p)$, the Tutte polynomial can be
written as : \begin{eqnarray}\label{eq1}
T(G,1+q\frac{(1-p)}p,\frac{1}{1-p})&=&\displaystyle\frac{(1-p)^{n-m-1}p^{1-n}}{q}\sum_{A\subset
E}p^{|A|}(1-p)^{m-|A|}q^{k(G,A)}\nonumber\\
&=&\displaystyle\frac{T(G,1/p,1/(1-p))}qE(q^{k(G,p)})\end{eqnarray}
where $E(q^k)$ denote the expectation of the random variable $q^k$
following the binomial distribution. This relation is well known because it gives a connexion between the Tutte polynomial and the partition function of the Potts model using the Fortuin Kasteleyn representation.

By using this calcul, we give
now a new result between the number of connected component for a
graph $G$ and its dual.
\section{Number of cluster per bond}
\subsection{ Definition and notations}
We associate with each edge of the graph $G$ a
distribution~: an edge will be white with probability $p$ and black
with probability $1-p$. We denote by $k(G,A,p)$ the number of
connected components with white edges in $A$, subset of $E$.\\

For the dual graph, we can build a relation between $A$ and a set
$A^{\star}$, subset of $E^{\star}$, the set of the edges of $G^{\star}$ : let be $e$ an edge of
$A$. These edge is common on two faces of the graph $G$ which
correspond two vertices of $G^{\star}$ and these two vertices define
an edge $e^{\star}$ in $G^{\star}$. So, given a subset $A$ of edges
$e$ of $E$, the set $A^{\star}$ of $E^{\star}$ is the complementary set of
edges $e^{\star}$ thus defined. In opposite, the edge in $G^{\star}$
shall be white with probability $1-p$ and black with probability
$p$. Also, if the edge $e$ in $G$ is white, the edge $e^{\star}$
will be black in $G^{\star}$. We denote by
$k(G^{\star},A^{\star},1-p)$ the number of connected components with
white edges in $A^{\star}$, subset of $E^{\star}$ (see Figure 2).\\
\begin{figure}[h!]
\begin{center}
\label{2carreA}
\includegraphics[width=3cm]{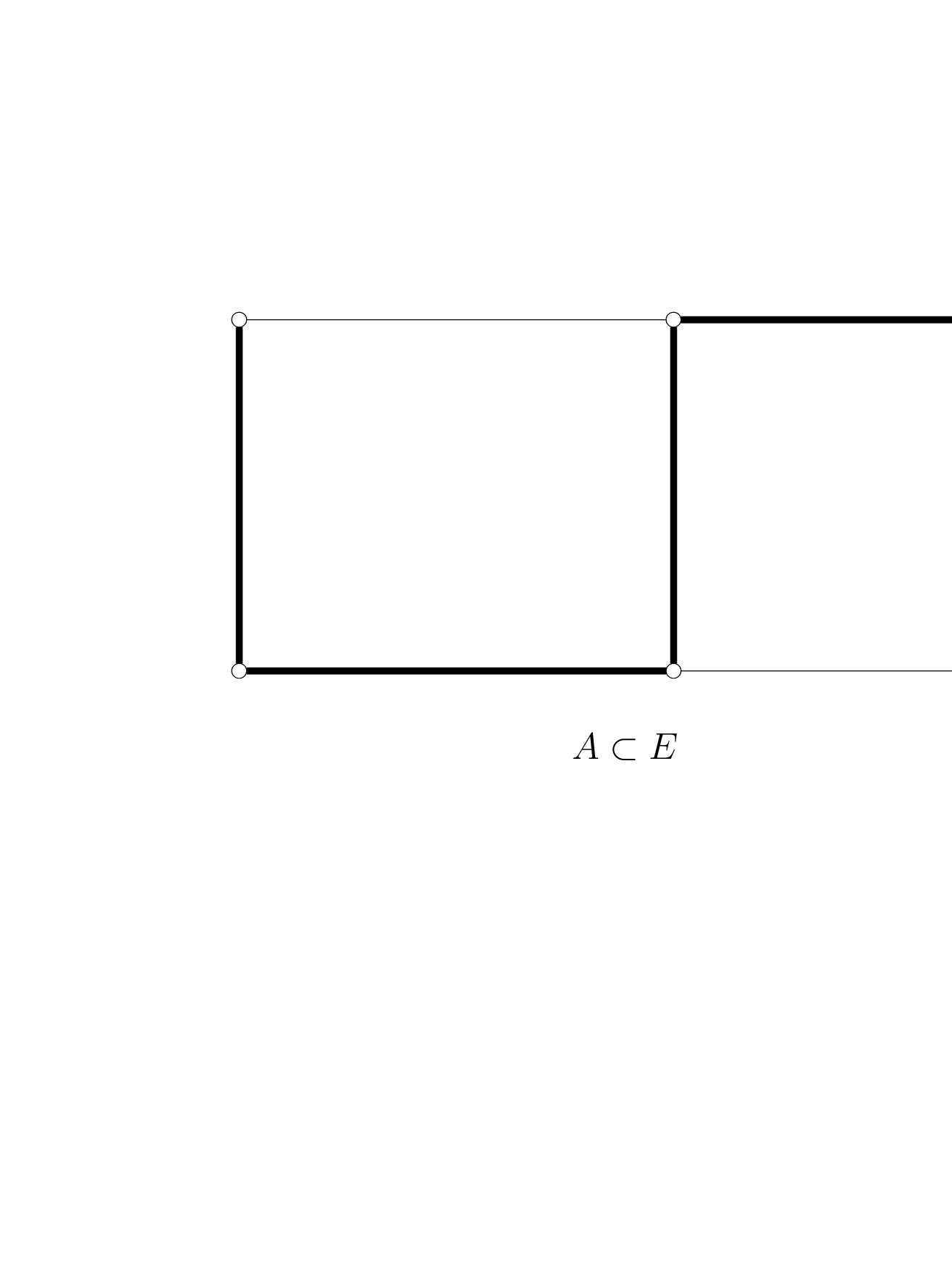}
\includegraphics[width=3cm]{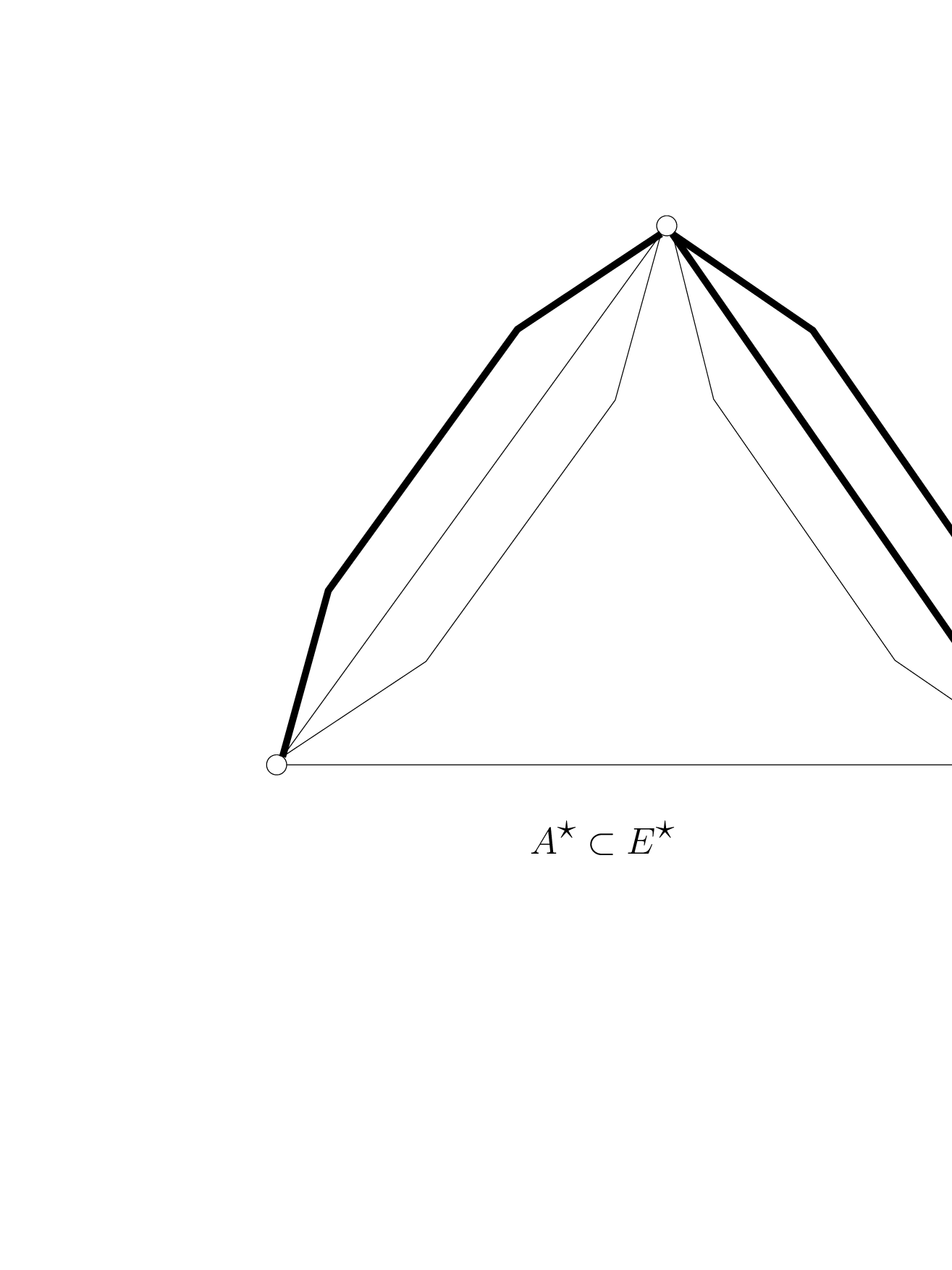}
\caption{\footnotesize Example of set $A$ in $G$ and corresponding ${A^\star}$ in $G^{\star}$.}
\end{center}
\end{figure}
\subsection {Connected components}
\noindent We use the following theorem and introduce a new proof based on the Tutte polynomial. It seems that this relation already appears in \cite{King02} in order to obtain some correlation duality relations for Potts model on planar graphs. 
\begin{teo}
Let be $G$ a connected planar graph and denote by $G^{\star}$ the
dual
graph of $G$.\\
For all subset $A$ of $E$ and corresponding subset $A^{\star}$ of
$E^{\star}$,
\begin{equation}\label{teo1}
k(G^{\star},A^{\star},1-p)-k(G,A,p)=|A|-n+1 \end{equation} where
$|A|$ represents number of elements of $A$.
\end{teo}
In this result, we can notice that the random variable $k(G^{\star},1-p) - k(G,p)$ follows a binomial distribution with modality set $\{1-n,\ldots,n^{\star}-1\}$, which is symmetric for $n=n^{\star}$.\\

\noindent{\bf Proof} : the proof is based on the symmetry of the
Tutte polynomial. For the graph $G$, on the curve $(x-1)(y-1)=q$,
the equation (\ref{eq1}) gives
$$\begin{array}{lll}
T(G,1+q\displaystyle\frac{1-p}p,\frac1{1-p})&=&(1-p)^{n-m-1}p^{1-n}\displaystyle\sum_{A\subset
e(G)}p^{|A|}(1-p)^{n_A-|A|}
q^{k(G,A,p)-1}\\
\end{array}$$
and, for the dual graph,
\begin{eqnarray*}
T(G^{\ast},1+q\displaystyle\frac{p}{1-p},\frac1{p})=p^{n^{\ast}-m^{\ast}-1}(1-p)^{1-n^{\ast}}\displaystyle\sum_{A^{\ast}\subset
e(G^{\star})}p^{n_A^{\ast}
-|A^{\ast}|}(1-p)^{|A^{\ast}|}q^{k(G^{\ast},A^{\ast},1-p)-1}.
\end{eqnarray*}
Notice that  $m^{\star}=m$, $m^{\star}-n^{\star}+1=f^{\star}=n-1$
and $1-n^{\star}=-f=-m+n-1$, so that the previous equation can be
written as
\begin{equation}\label{eq3}
T(G^{\ast},1+q\displaystyle\frac{p}{1-p},\frac1{p})=p^{1-n}(1-p)^{n-1}\displaystyle\sum_{A^{\ast}\subset
e(G^{\star})}p^{m-|A^{\ast}|}(1-p)^{|A^{\ast}|-m}q^{k(G^{\ast},A^{\ast},1-p)-1}.
\end{equation}
Moreover, the following relation $T(G^{\ast},x,y)=T(G,y,x)$ gives us
an other way to calculate (\ref{eq3}), namely~:\\

\noindent$T(G^{\ast},1+q\frac{p}{1-p},\frac1{p})= T(G,\frac{1}p,1+q\frac{p}{1-p})$\\
$\hspace*{2cm}= \displaystyle\frac{(1+q\frac{p}{1-p})^{m}}{\frac{1-p}{p}\left(q\frac{p}{1-p}\right)^{n}}\displaystyle\sum_{A\subset e(G)}
\left(\frac{q\frac{p}{1-p}}{1+q\frac{p}{1-p}}\right)^{|A|}\left(\frac{1}{1+q\frac{p}{1-p}}\right)^{m-|A|}q^{k(G,A,p)}.$\\

\noindent It becomes
\begin{equation}\label{eq4}
T(G^{\ast},1+q\displaystyle\frac{p}{1-p},\frac1{p})=p^{1-n}(1-p)^{n-1}\displaystyle\sum_{A\subset
e(G)}p^{|A|}(1-p)^{-|A|}q^{(k(G,A,p)+|A|-n)}
\end{equation}
Before, we made correspondence between the set $A$ and the set
$A^{\star}$, namely an edge $e$ ($e\in A$) is white in $G$ when the
corresponding edge $e^{\star}$ is black in $G^{\star}$. So, we
conclude $|A|+|A^{\ast}|=m$. $A^{\star}$ can be seen as the
complementary of $A$ in $G^{\star}$. In this way, the equation
(\ref{eq3}) gives us
\begin{eqnarray*}
T(G^{\ast},1+q\displaystyle\frac{p}{1-p},\frac1{p})=p^{1-n}(1-p)^{n-1}\displaystyle\sum_{A\subset
e(G)}p^{|A|}(1-p)^{-|A|}q^{k(G^{\ast},A^{\ast},1-p)-1}.
\end{eqnarray*}
Finally, if we compare this equality and the relation (\ref{eq3}),
it becomes
$$k(G^{\ast},A^{\ast},1-p)-k(G,A,p)=|A|-n+1.$$
\subsection{Mean of connected components}
The previous result yields a quick way to compare the means of
$k(G^{\ast},1-p)$ and $k(G,p)$ under the uniform distribution :
\begin{teo}\label{theo2}
$$\rm{E}(k(G^{\ast},1-p))-\rm{E}(k(G,p))=mp-n+1$$
\end{teo}
In fact, by multiplying equality (\ref{teo1}) by
$p^{|A|}(1-p)^{m-|A|}$ and by summing on all subsets $A$ of $E$,
we get :
$$\begin{array}{ll}
\hspace*{-1cm}\displaystyle\sum_{A\subset
e(G)}p^{|A|}(1-p)^{m-|A|}k(G^{\ast},A^{\ast},1-p)-\sum_{A\subset
e(G)}p^{|A|}(1-p)^{m-|A|}k(G,A,p)=&\\
&\hspace*{-5cm}\displaystyle\sum_{A\subset
e(G)}p^{|A|}(1-p)^{m-|A|}(|A|-n+1).
\end{array}$$
By the correspondence between the sets $A$ and $A^{\star}$ and the
relation $|A|+|A^{\star}|=m$, this equality becomes :
$$\begin{array}{ll}
\hspace*{-1cm}\displaystyle\sum_{A^{\star}\subset
e(G^{\star})}p^{m-|A^{\star}|}(1-p)^{|A^{\star}|}k(G^{\ast},A^{\ast},1-p)-\sum_{A\subset
e(G)}p^{|A|}(1-p)^{m-|A|}k(G,A,p)=&\\
&\hspace*{-5cm}\displaystyle\sum_{l=0}^mC_m^lp^{l}(1-p)^{m-l}(l-n+1).
\end{array}$$
which gives
$$\rm{E}(k(G^{\ast},1-p))-\rm{E}(k(G,p))=mp-n+1.$$
\noindent{\bf Some remarks :} (1) From the theorem \ref{theo2}, we
can notice the particular status played by the value $p=(n-1)/m$. For
this value, the two means are equal. We'll discuss on this value in
the next section for different graphs.\\
\noindent (2) Following \cite{Fill05}, we
can compare the expected values of the length of the minimal
spanning tree of $G$ denoted by $E(L_{MST}(G))$ and of $G^{\star}$
denoted by $E(L_{MST}(G^{\star}))$. They prove the following
relation $$E(L_{MST}(G))=\int_{0}^1E(k(G,p))\,dp-1.$$ In our
context, we obtain
$$\rm{E_{1-p}}(L_{MST}(G^{\star}))-\rm{E_p}(L_{MST}(G))=\frac{m}2-n+1.$$
In this equality, it appears the value $m=2(n-1)$. This relation
between number of edges and vertices of $G$ is a necessary condition
so that the Tutte polynomial may be symmetric in $x$ and $y$.
\subsection{ An other approach }
The result of theorem \ref{theo2} can be presented under another form : if we go back to the
equality (\ref{eq1}), we infer that
$$E(q^{k(G,p)})=\frac{qT(G,1+q\frac{(1-p)}p,\frac{1}{1-p})}{T(G,1/p,1/(1-p))}$$
and
$$E(q^{k(G^{\star},1-p)})=\frac{qT(G^{\star},1+q\frac{p}{1-p},\frac{1}{p})}{T(G,1/p,1/(1-p))}=
\frac{qT(G,\frac{1}{p},1+q\frac{p}{1-p})}{T(G,1/p,1/(1-p))}.$$ These
two expressions are differentiable in q so that
$$E(k(G,p)q^{k(G,p)-1})=\frac{T(G,1+q\frac{(1-p)}p,\frac{1}{1-p})}{T(G,1/p,1/(1-p))}+
q\frac{(1-p)}p\frac{\displaystyle\frac{\partial}{\partial x}T(G,1+q\frac{(1-p)}p,\frac{1}{1-p})}{T(G,1/p,1/(1-p))}$$
and
$$E(k(G^{\star},1-p)q^{k(G^{\star},1-p)-1})=\frac{T(G,\frac{1}{p},
1+q\frac{p}{1-p})}{T(G,1/p,1/(1-p))}+q\frac{p}{1-p}\frac{\displaystyle\frac{\partial}{\partial y}T(G,\frac{1}{p},1+q\frac{p}{1-p})}{T(G,1/p,1/(1-p))}.$$
Takinq 1 for $q$, we get :
$$E(k(G,p))-E(k(G^{\star},1-p))=p(1-p)\frac{\left[\frac{1}{p^2}\displaystyle\frac{\partial}{\partial x}T(G,\frac{1}{p},\frac{1}{1-p})
-\frac{1}{(1-p)^2}\displaystyle\frac{\partial}{\partial y}T(G,\frac{1}{p},\frac{1}{1-p})\right]}{T(G,1/p,1/(1-p))}$$
which can be written as
\begin{pro}
\begin{equation}
\label{eq2}
E(k(G,p))-E(k(G^{\star},1-p))=-p(1-p)\frac{d}{dp}\log T(G,\frac{1}p,\frac{1}{1-p}).\end{equation}
\end{pro}
By using the value
(\ref{p(1-p)}), we recover the relation of the theorem.\\

\noindent {\bf Remarks :\\}

1) For self dual graphs ($G=G^{\star}$), it is natural to take $p=1/2$ in (\ref{eq2}). For this value the right term in (\ref{eq2}) vanishes, thus $p=1/2$ can be seen as a minimum of the Tutte polynomial $T(G,x,y)$ on  the hyperbola $(x-1)(y-1)=q=1$ with $x=1/p$ and $y=1/(1-p)$. We can see that $ p=1/2= \sqrt q/(1+\sqrt q)$ for $q=1$. For other real values of $q$, we have a similar result that $p=\sqrt q/(1+\sqrt q)$ is a minimum of the Tutte polynomial on  the hyperbola $(x-1)(y-1)=q$. Notice that this particular value is conjectured to be the critical probability for q-Potts model.

In fact the Tutte polynomial for self dual graphs is symmetric and can be written as
$$
T(G,x,y)=\sum_{\alpha=1}^{n-1}\sum_{\beta=0}^{\alpha}a_{\alpha,\beta}(xy)^{\beta}(x^{\alpha-\beta}+ {y}^{\alpha-\beta}) 
$$
where $n$ is the number of vertices of $G$ and $a_{\alpha,\beta}$ are the coefficients of $T(G,x,y)$. On the hyperbola $(x=1+q/v,\,y=1+v)$,
$$\frac{dx}{dv}=-q/v^{2},\frac{dy}{dv}=1,\frac{d^2x}{dv^2}=2q/v^{3},\frac{d^2y}{dv^2}=0.$$
We can always write $T(G,x,y)=\displaystyle\sum_{\alpha=0}^{n_{s}-1}c_{\alpha}(x+y)^{\alpha}=T(G,v)$ thus $\displaystyle\frac{d}{dv}T(G,v)$ can be factorized by $1-q/v^{2}$ and vanishes for $v^{2}=q$ which corresponds to $x=y$.  By showing that the Tutte polynomial is a convex function of $v$, $v\geq 0$ otherwise $p\in]0,1[$ , $v=\sqrt q $ reaches the minimum and is equal to $T(G,1+\sqrt q, 1+\sqrt q)$.\\

\noindent 2) From the equation (\ref{eq1}), we obtain
$$
E(q^{k(G,p)-1}) = \frac{T(G,1+q\frac{(1-p)}p,\frac{1}{1-p})}{T(G,1/p,1/(1-p))}.
$$
By $i$ differentiations with respect to $q$, and taking $q=1$, the factorial moments are
\begin{equation}
\forall i\in\{1,\ldots,n\}, \, E[(k-1)(k-2)\ldots(k-i)] = ( \displaystyle\frac{1-p}{p}\Big) ^i \displaystyle\frac{\displaystyle\frac{\partial^i}{\partial x^i} T(G,\frac 1p,\frac{1}{1-p})}{T(G,\frac 1p,\frac{1}{1-p})}
\end{equation}
where $\displaystyle\frac{\partial^i}{\partial x^i} T$ is the $i$th partial derivative with respect to $x$ of Tutte polynomial.

\section{Applications}
\subsection{Finite planar graphs}
As example of finite planar graphs, we study the wheel which is a self dual graph. The wheel is the graph obtained from a cycle by adding a new vertex adjacent to all vertices in the cycle. This graph contains $n$ vertices and $2(n-1)$ edges. In this case, the equation in theorem \ref{theo2} can be written as follows
\begin{equation}
E\left[ k(G^{\star},1-p)\right]  - E\left[ k(G,p)\right] = 2(n-1)(p-\frac12) 
\end{equation}

The natural choice of $p$ is $0.5$ and so that for all graphs with $m=2(n-1)$. This condition is necessary for having a symmetric Tutte polynomial and in particular this is satisfied for self dual graphs.\\

Now, by looking at the result of the theorem \ref{theo2} and the previous example, we are led to study the particular value $p_{e}=\frac{n-1}{m}$ for general finite planar graphs. It comes $$\sum_{i=1}^{n} E( |C_{i}|^{-1})=E(k(G,p_{e}))=E(k(G^{\star},1-p_{e}))$$
where $C_{i}$ is the cluster containing the site $i$.\\

\begin{table}
\label{graph}
\begin{tabular}{|c|>{\centering}m{4.5cm}|c|c|c|c|}
\hline
Name & Graph & $|E|$ & $|V|$ & $p_e=\frac{|V|-1}{|E|}$ & $\displaystyle \lim_{s\to+\infty} p_e$ \\
\hline
$s$ squares & \includegraphics[width=3cm]{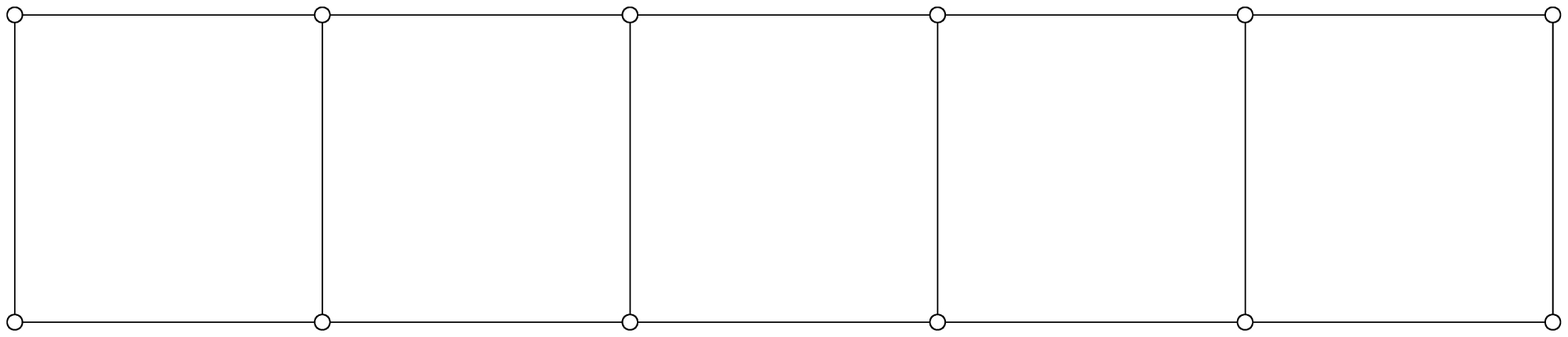} & $3s+1$ &$2s+2$  & $\displaystyle\frac{2s+1}{3s+1}$ & $2/3$\\
\hline
$s$ triangles & \includegraphics[width=3cm]{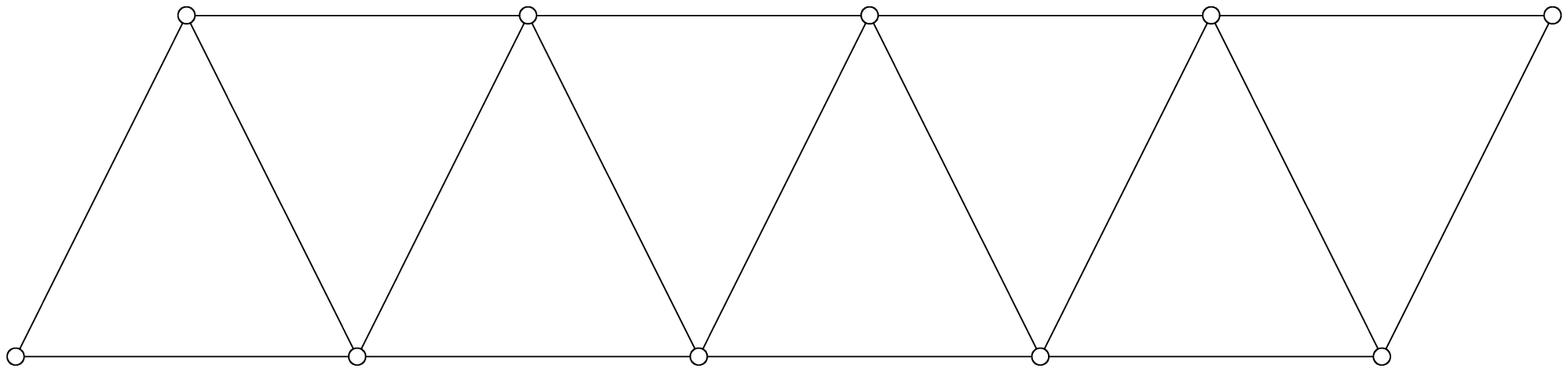} & $2s+1$ & $s+2$ & $\displaystyle\frac{s+1}{2s+1}$ & $1/2$\\
\hline
Square Lattice & \includegraphics[width=3cm]{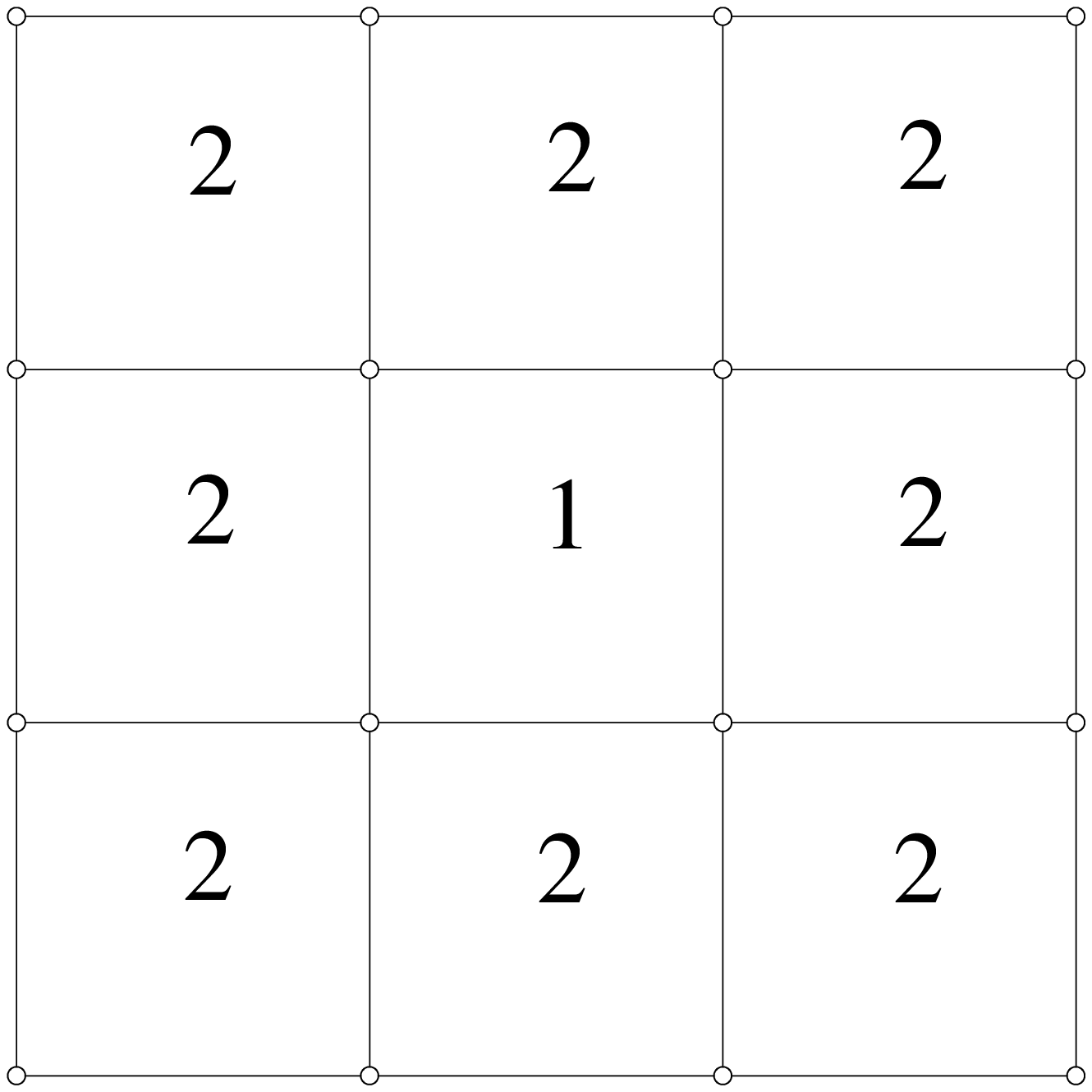} & $4s(2s-1)$ & $(2s)^2$ & $\displaystyle\frac{2s+1}{4s}$ & $1/2$\\
\hline
Bow-tie Lattice & \includegraphics[width=2cm]{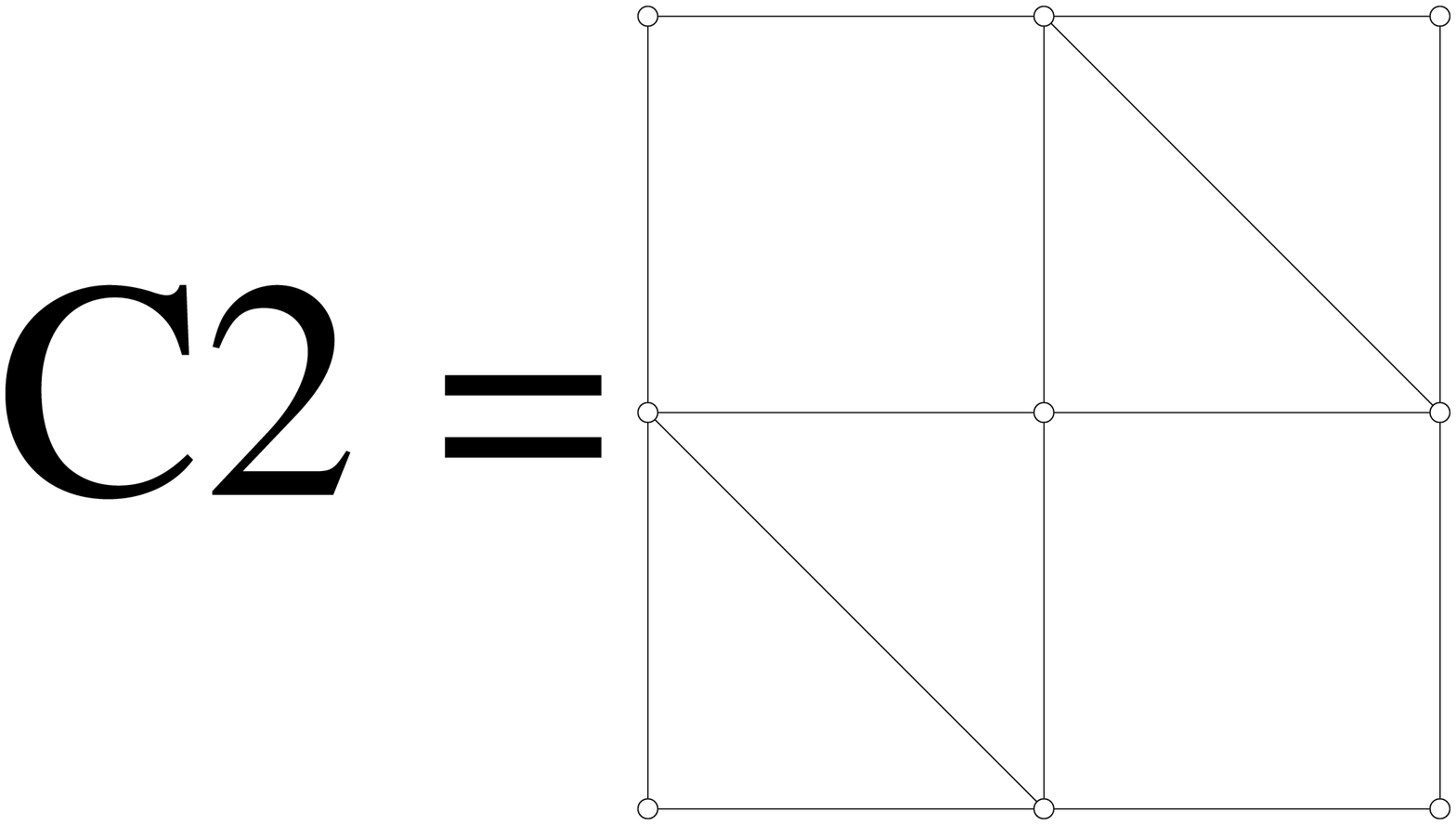} & $(4s-2)(10s-3)$ & $(4s-1)^2$ & $\displaystyle\frac{4s}{10s-3}$ & $2/5$\\
\hline
Crossed Matching & \includegraphics[width=2cm]{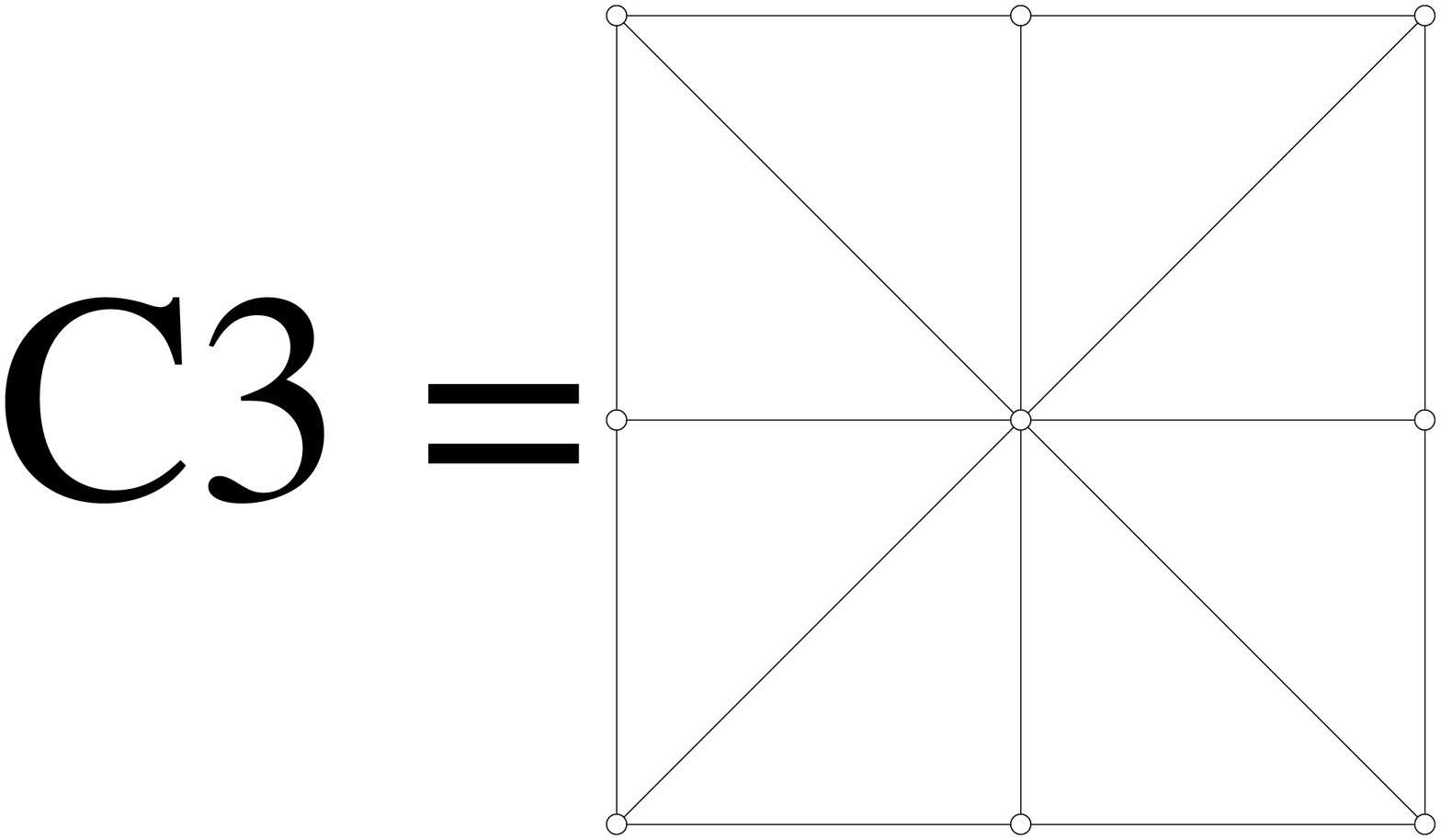} & $4(4s-2)(3s-1)$ & $(4s-1)^2$ & $\displaystyle\frac{s}{3s-1}$ & $1/3$\\
\hline
Octagonal & \includegraphics[width=2cm]{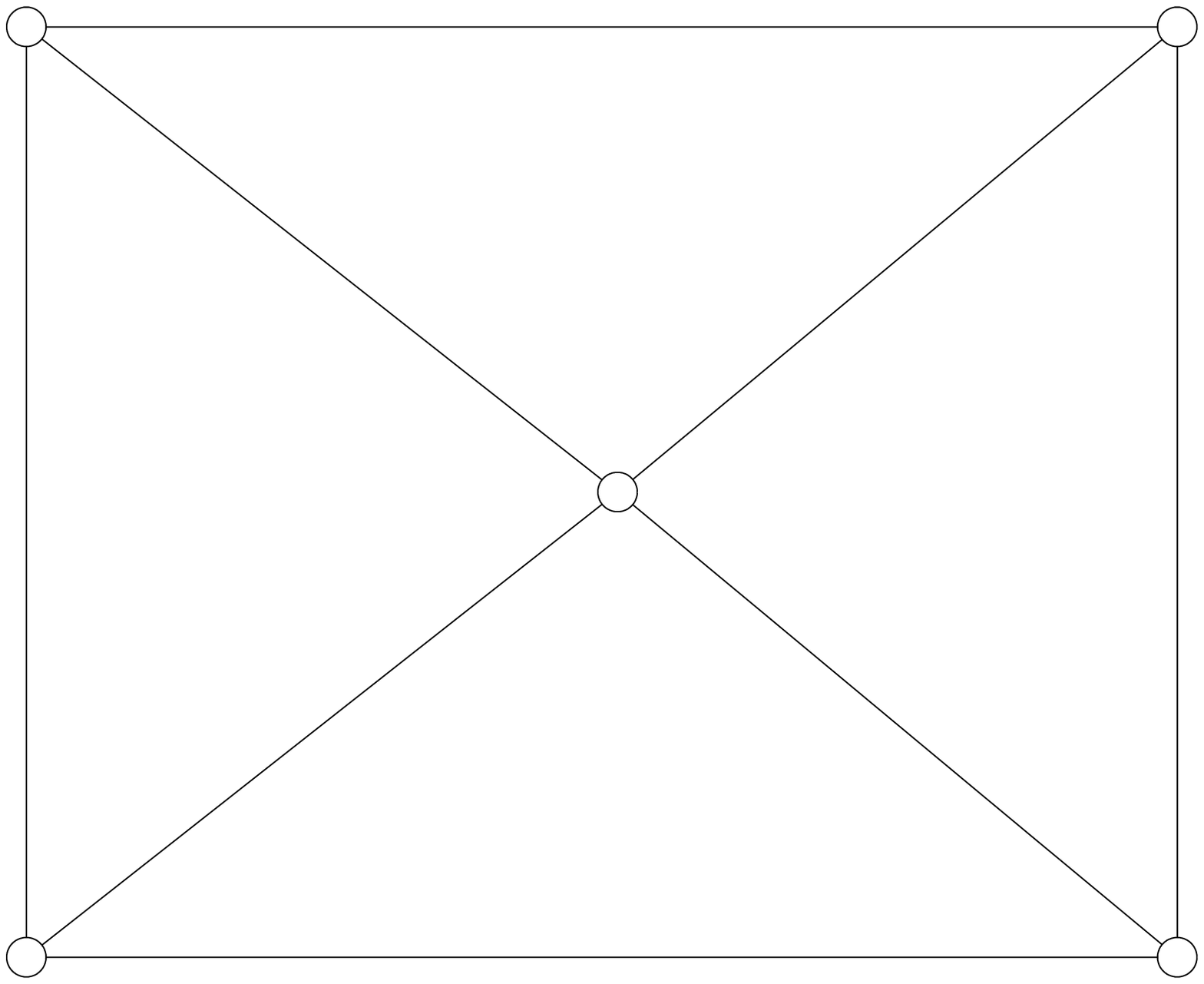} & $4(2s-1)(3s-1)$ & $8s^2-4s+1$ & $\displaystyle\frac{s}{3s-1}$ & $1/3$\\
\hline
Honeycomb Lattice & \includegraphics[width=3cm]{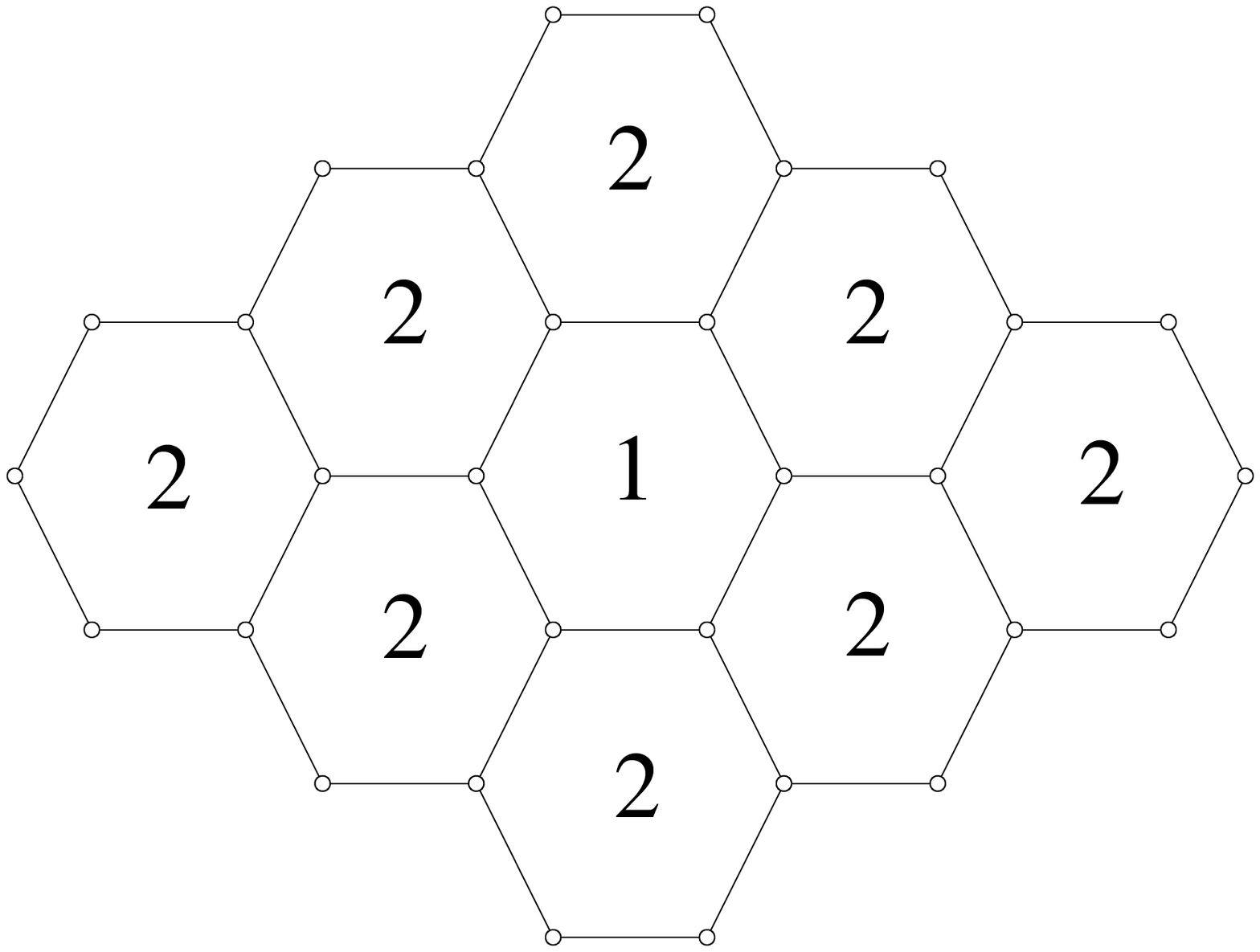} & $12s^2-4s-2$ & $8s^2-2$ & $\displaystyle\frac{8s^2-3}{12s^2-4s-2}$ & $2/3$\\
\hline
kagome lattice & \includegraphics[width=4cm]{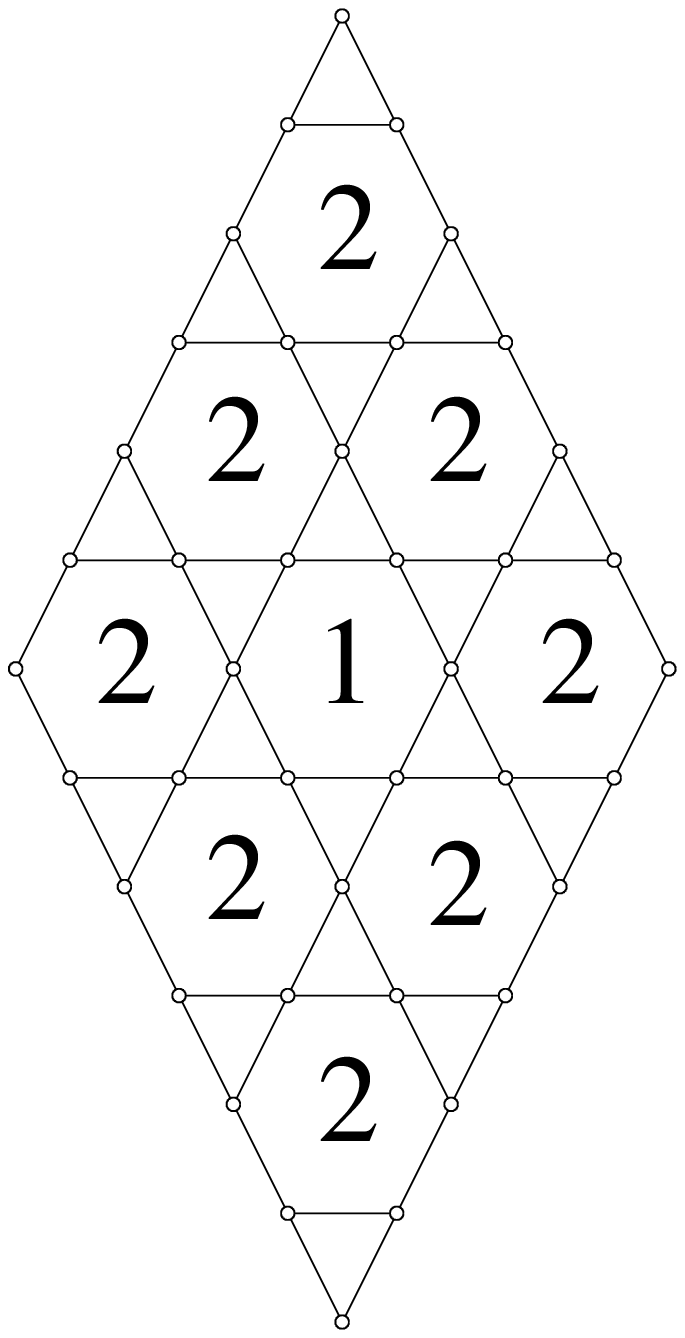} & $2(2s-1)(6s-1)$ & $4s(3s-1)$ & $\displaystyle\frac{(6s+1)}{2(6s-1)}$ & $1/2$\\
\hline
\end{tabular}
\caption{\footnotesize Some graphs with their particular value $p_e$.}
\end{table}

In table \ref{graph}, we present some graphs with their particular value $p_e$. For two firsts graphs, $s$ denotes respectively the number of squares and triangles. The third graph is the square lattice at step $s=2$. For the three following graphs, we just give the pattern associated with the square lattice, respectively : bow-tie lattice, crossed matching and octagonal. The two last graphs are respectively the honeycomb lattice and the kagome lattice at step $s=2$. For all graphs, we compute, at each step $s$, the number of edges and vertices and the corresponding value $p_e$.\\

\noindent Remark that $$\displaystyle p_{e}=\frac{n-1}{m}=\frac{2(n-1)}{\sum_{i=1}^{n}q_{i}}$$ where $q_{i}$ denotes the number of neighbours of the site $i$. 

\subsection{Infinite planar graphs}

Taking the thermodynamic limit in $\displaystyle p_{e}$, we obtain $p_{l}=\frac{2}{\bar q}$ where $\bar q$ is the coordination number of the graph. Obviously from Euler relation $$p_{e}+p_{e}^* =\frac{n-1}{m}+\frac{n^{\star}-1}{m}=1.$$
Of course, for self dual graph, $p_{l}=0.5$ : it corresponds to the critical bond percolation threshold of the square lattice.\\

In order to understand the last column of the table \ref{graph}, we write down the equation of theorem \ref{theo2} as follows
$$\rm{E}(k(G^{\ast},1-p_e))-\rm{E}(k(G,p_e))=mp_e-n+1.$$
By taking $p_e=\frac{n-1}{m}$, we obtain $$\rm{E}(k(G^{\ast},1-\frac{n-1}m))=\rm{E}(k(G,\frac{n-1}m)).$$
When $n$ tends to infinity with $m$, $m$ function of $n$, it becomes 
$$\rm{E}(k(G^{\star},1-\lim_{n\to+\infty} p_e))=\rm{E}(k(G,\lim_{n\to+\infty} p_e).$$
The last column of the table \ref{graph} gives value of these limits for our particular graphs. For honeycomb lattice $G$ and its dual, the triangular lattice, we find $$E(k(G,2/3))=E(k(G^{\star},1/3)).$$
How to understand for example this $2/3$ associated with honeycomb lattice~?
We can break up the expression of $p_{e}$ as $$p_{e}=\displaystyle\frac{2(n-1)}{\displaystyle\sum_{i=1}^{n_{int}}q_{i}+ \displaystyle\sum_{i=1}^{n_{ext}}q_{i} } = \displaystyle\frac{2}{ \displaystyle\frac{1}{n-1}\displaystyle\sum_{i=1}^{n_{int}}q_{i}+ \frac{1}{n-1}\displaystyle\sum_{i=1}^{n_{ext}}q_{i}}$$ with $n_{int}+n_{ext}=n$.  $n_{ext}$ is the number of vertices of the convex hull associated with honeycomb lattice at step $s$.
In this case and for all regular lattice with bounded degree, $\lim_{n\to+\infty}\displaystyle \frac{n_{ext}}{n}=0$, which gives $$\lim_{n\to+\infty}p_{e}=\lim_{n\to+\infty}\displaystyle\frac{2(n-1)}{\displaystyle\sum_{i=1}^{n_{int}}q_{i}}=\frac{2}{\bar q}.$$
So that to calculate this limit, we avoid  boundary effects. We only need of the coordination number of the graph : $3$ for the honeycomb lattice.
So we can state for Archimedean and Laves lattices the following relation $$E(k(G,\frac{2}{\bar q}))=E(k(G^{\star},1-\frac{2}{\bar q})).$$
\noindent{\bf Concluding remarks }
\begin{itemize}
\item[1)]The arithmetic and harmonic means of the number of neighbours of each vertex of the graph are the same for Archimedean lattices and also for Laves lattices with different weights as explained in \cite{Alm03}.
In this paper, he builts a new mean with other weights in order to classify  graphs with different bond percolation thresholds. Otherwise, results about ordering critical site and bond values with different connective constant are presented for some Archimedean lattices  in \cite{Wierman02}. 
\item[2)]In the preceding decomposition of ${\bar q}$, one may ask  what is the role of part concerning the vertices of the convex hull, it remembers of course conjecture connecting isoperimetric constant to percolation threshold of a graph. For example, it could be instructive to compare the speed of convergence of $p_e$ to $\frac{2}{\bar q}$ for different graphs.
\item[3)]Chang and Shrock \cite{Chang04} remark that for strips of graphs such as square, triangular, kagome, that complex singularity of the number of cluters per bond are precisely on the boundary of the accumulation set of the zeros of the partition function.
\item[4)]The present study shows the major role of self-dual graphs. An interesting perspective will be the location of complex zeros of their Tutte polynomial and a discussion on the thermodynamic questions.

\end{itemize}

\bibliographystyle{plain}

\bibliography{eb,horst,biblio,markov,env,bbd,tutte}

\end{document}